\documentclass[showkeys,twocolumn,amsmath,amssymb,aps,pra]{revtex4-2}
\usepackage{bm}
\usepackage{amsfonts}
\usepackage{amsmath}
\usepackage{graphicx}
\usepackage{float}

\begin{document} 
\title{A concise and universal method for deriving arbitrary paraxial and d'Alembertian cylindrical Gaussian-type light modes}
\author{Tomasz Rado\.zycki}
\email{t.radozycki@uksw.edu.pl}
\affiliation{Faculty of Mathematics and Natural Sciences, College of Sciences, Institute of Physical Sciences, Cardinal Stefan Wyszy\'nski University, W\'oycickiego 1/3, 01-938 Warsaw, Poland} 

\begin{abstract}
A concise method of deriving expressions for Gaussian-like solutions of the paraxial and d'Alembert equations is presented. This method is based on the Hankel transform. Choosing some Gaussian base functions with slight modifications of the prefactor all basic beams of cylindrical character can be easily obtained. This refers to Gaussian, Bessel-Gaussian, modified Bessel-Gaussian, Laguerre-Gaussian and Kummer-Gaussian (i.e., Hypergeometric-Gaussian) beams although potentially other beams can come into play as well. For instance a new type of a beam that can be derived in this way is described through the incomplete gamma function so it may be called a $\gamma$ beam.
\end{abstract}

\maketitle

\section{Introduction}\label{intr}

Over the past few decades, a number of Gaussian-like beams have been obtained in the mathematical sense, i.e., as solutions to the paraxial, Helmholtz or Maxwell equations, and later experimentally realized. It is a rapidly expanding area of science due to its numerous and significant practical applications ranging from optical trapping and guiding through image processing, optical communication, harmonics generation, quantum cryptography to biology and medicine~\cite{ste,fazal,pad,woe,bowpa,grier1,kol,alt,nis,cc}. One can mention here pure Gaussian beams~\cite{davis,sie,nemo,mw,saleh,ibbz,sesh,gustavo,er,selina} with and without vortex component, Bessel-Gaussian (BG)~\cite{saleh,gori,april1,mendoza} and Laguerre-Gaussian (LG) beams~\cite{sie,saleh,mendoza,lg,lg2,april2,april3,nas} of various orders, parabolic and Mathieu beams~\cite{fb,fa} or Kummer-Gaussian (KG) (i.e.,  Hypergeometric-Gaussian) beams~\cite{kot,karimi}. Viewed from a purely theoretical perspective, these mathematical expressions describing the aforementioned beams were derived directly by solving wave equations~\cite{davis,sesh,er,selina,jordan,hall,ibbz}, using the Fresnel diffraction formula~\cite{gori,goo,zamboni}, via superposing Gaussian beams, Bessel beams or even plane waves~\cite{ibbz,wen,ding,bagini} or exploiting various forms of the Schr\"odinger propagator in $2D$~\cite{tr1,tr2}. 

Several attempts have been made in the past to unify both the derivation and the description of the paraxial beams~\cite{end,li,vl,fe}. Somewhat surprisingly, to the best of our knowledge, the systematic approach involving the Hankel transform~\cite{patra,erde}, which is natural for problems with cylindrical symmetry, has not been implemented so far.
In the present work we would like to fill this gap and propose a simple and universal method, which allows treating all those beams on an equal footing. Moreover, it reveals a very simple link between paraxial beams and those described by the d'Alembert equation. In place of performing the paraxial approximation in the latter, a simple substitution is sufficient. One can say that in this way the description of both full and approximate beams is unified as well. 

Apart from this pleasant feature, this approach can provide seemingly a starting point for generating other modes that satisfy the paraxial, Helmholtz or d'Alembert equations, by appropriately choosing the function $g_n(s)$ defined through Eq.~(\ref {phit2}). As an example of this potential application the so called $\gamma$ beam is briefly discussed in Sec.~\ref{gammab}. 

The method in the present form is developed within the scalar approximation. However, from the d'Alembert solutions the exact vector wave satisfying Maxwell equations can be restored via the construction of~\cite{ibbz} using the Riemann-Silberstein vector ${\mathbf F}= {\mathbf E}+ic{\mathbf B}$. Then the scalar functions found in this work can serve as the corresponding Whittaker potentials~\cite{whi}. 

Naturally, the Hankel transform exploits the cylindrical coordinates and, therefore, the method is not suited for non-cylindrical beams like Hermite-Gaussian or  Ince-Gaussian beams. 

In the following section the details of this method are formulated, first for paraxial and then for d'Alembert modes and in Sec.~\ref{ebi} it is applied to derive the particular beams spoken of above. In the last section it is shown, that the paraxial approximation of the full equation practically reduces to the simple substitution $\alpha(\zeta)\longmapsto\alpha(z)$, which enables the collective treatment of all beams and thus clarifies the results obtained in Sec.~\ref{hankeltrans}. 

\section{Hankel-transform method}\label{hankeltrans}

\subsection{Monochromatic paraxial beams}\label{mp}

A monochromatic light wave of frequency $\omega$ propagating along the $z$-axis can be represented through the complex electric field with a scalar function $\Psi(\bm{r},z,t)$ as follows:
\begin{equation}
\bm{E}(\bm{r},z,t)=\bm{E}_0\Psi(\bm{r},z,t),
\label{ep}
\end{equation}
where $\bm{r}$ stands for transverse coordinates only. In the present work the focus is put on the properties of  $\Psi$, without referring to the polarization, so $\bm{E}_0$ can be chosen for instance 
as a constant vector. The electric field~(\ref{ep}), within the standard scalar approximation \mbox{[$\bm {\nabla}(\bm{\nabla E})\longmapsto 0$]}, satisfies the d'Alembert equation.  In the paraxial regime it can be further reduced by first substituting
\begin{equation}
\Psi(\bm{r},z,t)=e^{ik(z-c t)}\psi(\bm{r},z),
\label{epi}
\end{equation}
and next neglecting the second order derivative with respect to $z$ with the assumption
\begin{equation}
|\partial^2_z\psi|\ll |k\partial_z\psi|,
\label{parap}
\end{equation}  
where $\partial_z$ denotes $\partial/\partial_z$. This yields the well-known equation, called the {\em paraxial equation} for the envelope~\cite{sie}:
\begin{equation}\label{paraxial}
\mathcal{4}_\perp \psi(\bm{r},z)+2ik\partial_z \psi({\bm{r}},z)=0.
\end{equation}
When deriving~(\ref{paraxial}) the $z$-dependence is separated so the Laplace operator $\mathcal{4}_\perp$ is two-dimensional including differentiations over transverse coordinates only. 

Further simplification is achieved with the use of the polar coordinates $r=\sqrt{x^2+y^2}$ and $\varphi$, which allows for isolating the dependence on the angular variable by means of the substitution
\begin{equation}
\psi(r,\varphi,z)=e^{i n\varphi}\Phi(r,z),\;\;\;\mathrm{where}\,\;\; n\in\mathbb{Z}.
\label{phi}
\end{equation}
The paraxial equation is then given the form
\begin{equation}\label{parcylf}
\Big(\partial_r^2+\frac{1}{r}\,\partial_r-\frac{n^2}{r^2}+2ik\partial_z\Big)\Phi(r,z)=0,
\end{equation}
which constitutes the starting point for the proposed method.

Let us begin with performing on Eq.~(\ref{parcylf}) the $n$th-order Hankel transform with respect to the radial variable $\xi$, which is defined as~\cite{patra}
\begin{equation}
\hat{f}_n(s)=\int\limits_0^\infty \mathrm{d}x\, x J_n(sx)f(x),
\label{hankel}
\end{equation}
with $J_n(x)$ being the Bessel function of the first kind. It is well known that this procedure
leads to the simplified equation 
\begin{equation}
(2ik\partial_z-s^2)\hat{\Phi}_n(s,z)=0.
\label{phit1}
\end{equation}
The index $n$ does not indicate here that this function depends on the parameter $n$, but merely reminds that it is associated with $\Phi(r,z)$ through the transform of the $n$th order. This will become clear later.

Eq.~(\ref{phit1}) can now be integrated in a straightforward way, and one gets
\begin{equation}
\hat{\Phi}_n(s,z)=g_n(s)e^{-is^2z/2k},
\label{phit2}
\end{equation}
where the function $g_n(s)$ is arbitrary and constrained only by the requirement of the inverse transform to exist. This inverse transform has the form identical to~(\ref{hankel})~\cite{patra}:
\begin{equation}
f(x)=\int\limits_0^\infty \mathrm{d}s\, s J_n(xs)\hat{f}_n(s),
\label{hankelinv}
\end{equation}
so the solution of Eq.~(\ref{parcylf}) is obtained as
\begin{equation}
\psi(r,\varphi,z)=e^{in\varphi}\int\limits_0^\infty \mathrm{d}s\, s J_n(r s )g_n(s)e^{-is^2z/2k}.
\label{inp}
\end{equation}
This type of an equation has already appeared in the context of conical refraction~\cite{sok,turp,myl}. Any function of the form~(\ref{epi}) with the envelope $\psi(r,\varphi,z)$ as given above represents a certain light beam, as long as it is carrying finite energy per unit length. The choice of $g_n(s)$ is up to us and hence various light beams can be obtained from the above universal formula. To ensure the finite value of the transported energy, it will be assumed in this work that this function displays the Gaussian factor in the transverse plane, which limits the class of beams considered in this study to the Gaussian-type beams, although with an exception for a certain specific choice of $g_n(s)$ to be considered later.

Consequently the function $g_n(s)$ is chosen in the form:
\begin{equation}
g_n(s)=\beta_n(s)e^{-w_0^2s^2/4},
\label{gaussb}
\end{equation}
where $w_0$ will prove to be the beam waist. This exponential factor will be shared in all $g_n$'s below since the inverse Hankel transform is expected to turn it into the ordinary Gaussian factor:
\begin{equation}
e^{-w_0^2s^2/4}\;\;\longmapsto\;\; e^{-r^2/w_0^2}.
\label{trga}
\end{equation}
This conclusion, however, obviously depends on the complete form of $g_n(s)$ and one exception to it will be pointed out in Sec.~\ref{gammab}.  Therefore, it is the form of the {\em prefactor} $\beta_n(s)$ that is at our disposal and enables to generate various paraxial beams as desired, considered in Sec.~\ref{ebi}. 

With the use of~(\ref{gaussb}) our principal formula can be given the form
\begin{eqnarray}
\Psi_{\mathrm{p},k}&&(r,\varphi,z,t)=\label{inpf}\\
&&e^{ik(z-c t)}e^{in\varphi}\!\int\limits_0^\infty\! \mathrm{d}s\, s J_n(r s )\beta_n(s)e^{-\alpha(z)s^2/4},
\nonumber
\end{eqnarray}
where $\alpha(z)=w_0^2 +2iz/k$, the subscript $\mathrm{p}$ stands for ``paraxial'' and $k$ being a reminiscent of wave-number dependence. In the following subsection it is shown that the same formula can be used to obtain non-paraxial beams embodying similar characteristics.

\subsection{D'Alembert pulses}\label{npb}

Gaussian-type d'Alembert pulses called also focus wave modes~\cite{fig} are not monochromatic. In order to derive appropriate analytical formulas the full d'Alembert equation has to be dealt with:
\begin{equation}
\Big(\mathcal{4}_\perp +\partial^2_z-\frac{1}{c^2}\,\partial^2_t \Big)\Psi(\bm{r},z,t)=0,
\label{dal}
\end{equation}
but, as mentioned, the results of the previous section (with certain obvious redefinitions) are still applicable. 

Before one proceeds with the Hankel transform it is convenient to introduce two new variables~\cite{ibbz}:
\begin{equation}
\zeta=\frac{1}{2}(z+ct),\;\;\;\; \eta=z-ct,
\label{zeta}
\end{equation}
which allows to rewrite the d'Alembert equation in the form
\begin{equation}
\big(\mathcal{4}_\perp +2\partial_\zeta\partial_\eta \big)\Psi(\bm{r},\zeta,\eta)=0.
\label{dale}
\end{equation}
In the polar coordinates the $\varphi$ and $\eta$ dependences can be isolated by writing
\begin{eqnarray}
\Psi(r,\varphi,\zeta,\eta)&&\;=e^{ik\eta}\psi(r,\varphi,\zeta)\label{is}\\
 &&\;=e^{ik\eta}e^{i n\varphi}\Phi(r,\zeta)\;\;\;\;\mathrm{where}\;\;\; n\in\mathbb{Z},\nonumber
\end{eqnarray}
with $k$ being a constant, and the wave-equation takes the form
\begin{equation}
\Big(\partial_r^2+\frac{1}{r}\,\partial_r-\frac{n^2}{r^2}+2ik\partial_\zeta\Big)\Phi(r,\zeta)=0.
\label{dalu}
\end{equation}
This equation is practically identical to~(\ref{parcylf}) so all steps leading to~(\ref{inpf}) can be repeated, with the final result:
\begin{equation}
\Psi_{\mathrm{A},k}(r,\varphi,\zeta,\eta)=e^{ik\eta}e^{in\varphi}\int\limits_0^\infty \mathrm{d}s\, s J_n(r s )\beta_n(s)e^{-\alpha(\zeta)s^2/4},
\label{inpfx}
\end{equation}
where $A$ stands for ``d'Alembert'' and the role of the subscript $k$ is identical as in~(\ref{inpf}). When comparing this expression with the latter it becomes obvious that the formulas for non-paraxial beams can be easily obtained from the paraxial ones (or {\em vice versa}) with the trivial substitution $z\longmapsto\zeta$. Surely, the non-paraxial beam is not monochromatic, and apart from $\eta$ it depends on time through $\zeta$. Therefore, the interpretation of the parameter $k$ is different in both cases.

From the solutions of the d'Alembert equation in the form of~(\ref{inpfx}) the corresponding Helmholtz ($H$) beams can eventually be obtained. As mentioned above the role of the number $k$ is different in the case of the d'Alembert  equation so one cannot expect that the particular solution~(\ref{inpfx}) is directly transformed into the Helmholtz solution with $k$ representing its wave number $\omega/c$. Rather a superposition of the former solutions with a certain profile $D(k)$, which is assumed to be smooth and slowly varying function, i.e,
\begin{equation}
\Psi_{\mathrm{A}}(r,\varphi,\zeta,\eta)=\int\limits_{-\infty}^\infty \mathrm{d}k\, D(k)\Psi_{\mathrm{A},k}(r,\varphi,\zeta,\eta)
\label{supa},
\end{equation}
should be the starting point for this reduction. Then the monochromatic Helmholtz solution can be found via the following Fourier-type integration:
\begin{eqnarray}
\Psi_{\mathrm{H},\omega/c}&&(r,\varphi,z,t)=\label{helm}\\
&&e^{-i\omega t}\int\limits_{-\infty}^\infty \mathrm{d}t' e^{i\omega t'}\Psi_{\mathrm{A}}(r,\varphi,(z+ct')/2,z-ct'),
\nonumber
\end{eqnarray}
which is easy to execute since it reduces to the Dirac delta function due to the form of $\Psi_{\mathrm{A},k}$. This fact will be exploited later, in Sect.~\ref{parapp}.

\section{Application for various beams}\label{ebi}

In this section it will be demonstrated that practically all the expressions for cylindrical beams with a Gaussian profile, well known in optics, can be derived in a very easy manner starting from Eq.~(\ref{inpf}) and simply modyfying the factor $\beta_n(s)$. The following examples will be dealt with in turn: the Gaussian beam, the Bessel-Gaussian beam, the modified Bessel-Gaussian beam, the Laguerre-Gaussian beam, the so called $\gamma$-beam and finally the general Kummer-Gaussian beam.

\subsection{Gaussian beam}\label{gabi} 

The fundamental paraxial beam bearing the name of the Gaussian beam (of the $n$th order) can be generated from the expression~(\ref{inp}) by choosing
\begin{equation}
\beta_n(s)=s^n,
\label{betagauss}
\end{equation}
which includes also the simplest case $s^0=1$. The exact form of $\beta_n(s)$ will prove to play the crucial role as will be shown in this section. 

Substituting $\beta_n(s)$ in the form~(\ref{betagauss}) into~(\ref{inpf}) one obtains
\begin{equation}
\Psi_{\mathrm{p},k}(r,\varphi,z,t)=e^{ik(z-c t)}e^{in\varphi}\int\limits_0^\infty \mathrm{d}s\, s^{n+1} J_n(r s )e^{-\alpha(z)s^2/4},
\label{inpfg}
\end{equation}
and after performing the integration with respect to $s$, one arrives at~\cite{erde}
\begin{equation}
\psi(r,\varphi,z)=\Big(\frac{2}{\alpha(z)}\Big)^{n+1}r^ne^{in\varphi}e^{-r^2/\alpha(z)}.
\label{gbnp}
\end{equation}
For brevity the formulas for $\psi(r,\varphi,z)$ are given. The whole expressions for $\Psi_{\mathrm{p},k}(r,\varphi,z,t)$ are easily recovered by introducing the plane-wave phase factor as in~(\ref{epi}). The obtained formula represents just the Gaussian beam of order $n$, (i.e. bearing the vorticity index $n$, which can also vanish, if desired) expressed in the used notation. In order to give it the form that optical physicists are more accustomed to, the following identifications can be used:
\begin{subequations}\label{nota}
\begin{align}
\alpha(z)&=w_0^2+2i\,\frac{z}{k}=w_0^2\Big(1+i\,\frac{z}{z_R}\Big),\label{nota1}\\
\frac{1}{\alpha(z)}&=\frac{k}{2z_R}\,\frac{w_0}{w(z)}e^{-i\psi_G(z)},\label{nota2}
\end{align}
\end{subequations}
where $w_0$ is the beam waist, $z_R=kw_0^2/2$ denotes the Rayleigh length,  $w(z)=w_0\sqrt{1+(z/z_R)^2}$ is the beam radius, $R(z)=z(1+(z_R/z)^2)$ stands for the wavefront curvature and $\psi_G(z)=\arctan (z/z_R)$ is the  Gouy phase.
In this explicit notation the expression~(\ref{gbnp}) takes the well-known form:
\begin{eqnarray}
\Psi_{\mathrm{p},k}&&(r,\varphi,z,t)={\cal C}\Big(\frac{w_0}{w(z)}\Big)^{n+1}e^{ik(z-c t)}r^ne^{in\varphi}\label{gaussbeamn}\\
&&\times\exp\Big[-\frac{r^2}{w(z)^2}+i\frac{kr^2 }{2 R(z)}
-i(n+1)\psi_G(z)\Big],\nonumber
\end{eqnarray}
where the inessential constants have been absorbed into the normalization constant having been denoted with $\cal C$ (omitted in the subsequent subsections). 

The procedure leading to the non-paraxial Gaussian pulse is now extremely easy. In accordance with the formula~(\ref{inpfx}), it is enough to insert $z=\zeta$ into~(\ref{gaussbeamn}), with the following result:
\begin{eqnarray}
\Psi_{\mathrm{A},k}&&(r,\varphi,\zeta,\eta)={\cal C}\Big(\frac{w_0}{w(\zeta)}\Big)^{n+1}e^{ik\eta}r^ne^{in\varphi}\label{gaussbeamv}\\
&&\times\exp\Big[-\frac{r^2}{w(\zeta)^2}+i\frac{kr^2 }{2 R(\zeta)}
-i(n+1)\psi_G(\zeta)\Big].\nonumber
\end{eqnarray}
This satisfies the d'Alembert equation, which can be verified through the direct computation, and agrees with Ref.~\cite{ibbz}. Since $w(\zeta)\rightarrow \infty$ as $t\rightarrow\pm\infty$ and $z\rightarrow\pm\infty$ the exact expression~(\ref{gaussbeamv}) represents a diffracting pulse.

\subsection{Bessel-Gaussian beam}\label{begabi} 

If instead of $\beta_n(s)=s^n$ the prefactor in the form of modified (i.e. hyperbolic) Bessel function is chosen, i.e.
\begin{equation}
\beta_n(s)=I_n(\chi s),
\label{gaussbeb}
\end{equation}
the Bessel-Gaussian beam is obtained. The parameter $\chi$ is at our disposal and, as will be seen below, is related to the aperture angle of the beam. Hence
\begin{equation}
\psi(r,\varphi,z)=e^{i n\varphi}\int\limits_0^\infty \mathrm{d}s\, s J_n(r s )I_n(\chi s)e^{-\alpha(z)s^2/4},
\label{inpfbg}
\end{equation}
and after the integration over $s$ the envelope in the following form is obtained~\cite{erde}:
\begin{equation}
\psi(r,\varphi,z)=\frac{2}{\alpha(z)}\,e^{\chi^2/\alpha(z)}e^{in\varphi}J_n\Big(\frac{2\chi r}{\alpha(z)}\Big)e^{-r^2/\alpha(z)}.
\label{gaussbebeam}
\end{equation}
This expression is again perfectly known. It represents the Bessel-Gaussian beam of the $n$th order and may be given the traditional form upon applying the notation~(\ref{nota}) together with the identification $\chi\longmapsto z_R\sin\theta$~\cite{bor,mad}:
\begin{eqnarray}
&&\Psi_{\mathrm{p},k}(r,\varphi,z,t)=\frac{w_0}{w(z)}e^{ik(z-c t)}e^{in\varphi}J_n\Big(\frac{kr\sin\theta}{1+i\frac{z}{z_R}}\Big)\label{super}\\
&&\exp\Big[-\frac{r^2}{w(z)^2}+ \frac{ikr^2 }{2 R(z)}
-\frac{ikz}{2(1+i\frac{z}{z_R})}\sin^2\theta-i\psi_G(z)\Big],\nonumber
\end{eqnarray}
to be compared with~\cite{gori}.
As mentioned, the normalization constant is omitted. 

The non-paraxial Bessel-Gaussian pulse can now be obtained in the described way by the substitution $z\longmapsto \zeta$, and one gets
\begin{equation}
\Psi_{\mathrm{A},k}(r,\varphi,\zeta,\eta)=e^{ik\eta} \psi(r,\varphi,\zeta),
\label{pbgfnp}
\end{equation}
with $\psi$ defined in~(\ref{gaussbebeam}).

\subsection{Modified Bessel-Gaussian beam}\label{mobegabi} 

Since the Hankel transform with the additional Gaussian factor converts modified Bessel functions into Bessel functions of the first kind and {\em vice versa}~\cite{erde}, not surprisingly in order to obtain modified Bessel-Gaussian beam~\cite{bagini} one has to choose
\begin{equation}
\beta_n(s)=J_n(\chi s)
\label{mgaussbeb}
\end{equation}
instead of~(\ref{gaussbeb}). This leads to 
\begin{equation}
\psi(r,\varphi,z)=e^{i n\varphi}\int\limits_0^\infty \mathrm{d}s\, s J_n(r s )J_n(\chi s)e^{-\alpha(z)s^2/4},
\label{inpfmbg}
\end{equation}
and consequently after almost identical integration one obtains
\begin{equation}
\psi(r,\varphi,z)=\frac{2}{\alpha(z)}\,e^{-\chi^2/\alpha(z)}e^{in\varphi}I_n\Big(\frac{2\chi r}{\alpha(z)}\Big)e^{-r^2/\alpha(z)}.
\label{mgaussbebeam}
\end{equation}
 which accounts for the modified Bessel-Gaussian beam of the $n$th order, the profile of which contains the modified Bessel function. After applying~(\ref{nota}) together with $\chi= z_R\sin\theta$ the beam acquires its conventional form:~\cite{bagini}:
\begin{eqnarray}
&&\Psi_{\mathrm{p},k}(r,\varphi,z,t)=\frac{w_0}{w(z)}e^{ik(z-c t)}e^{in\varphi}I_n\Big(\frac{kr\sin\theta}{1+i\frac{z}{z_R}}\Big)\label{msuper}\\
&&\exp\Big[-\frac{r^2}{w(z)^2}+ \frac{ikr^2 }{2 R(z)}
+\frac{ikz}{2(1+i\frac{z}{z_R})}\sin^2\theta-i\psi_G(z)\Big],\nonumber
\end{eqnarray}

For the non-paraxial modified Bessel-Gaussian pulse the formula~(\ref{pbgfnp}) holds again.

\subsection{Elegant Laguerre-Gaussian beam}\label{lgb}

Slight modification of the prefactor $\beta_n(s)$ in~(\ref{betagauss}), namely
\begin{equation}
s^n\;\;\longmapsto\;\;s^{n+2p},\;\;\;\;\mathrm{where}\;\; p=0,1,2,\ldots
\label{lgs}
\end{equation}
leads to the so called elegant Laguerre-Gaussian beam. The calculation goes in a similar manner. First we substitute  
$\beta_n(s)$ in this form into the inverse Hankel transform obtaining
\begin{equation}
\psi(r,\varphi,z)=e^{i n\varphi}\int\limits_0^\infty \mathrm{d}s\, s^{n+2p+1} J_n(r s)e^{-\alpha(z)s^2/4},
\label{inpflg}
\end{equation}
and then after executing the integral the envelope is found in the form:
\begin{eqnarray}
\psi(r,\varphi,z)=&&\,2^pp!\, \frac{(n+p)_p}{(n+1)_p}\Big(\frac{2}{\alpha(z)}\Big)^{n+p+1}\!\!\! r^n e^{in\varphi}\nonumber\\
&&\, \times L_p^n\Big(\frac{r^2}{\alpha(z)}\Big)e^{-r^2/\alpha(z)},
\label{gausslbeam}
\end{eqnarray}
where $L_p^n(z)$ denote the associated Laguerre polynomials and $(n)_k$ stands for the Pochhammer symbol. The beam~(\ref{gausslbeam}) bears the name of the {\em elegant} LG beam with its standard description as~\cite{sie,april2,april3}:
\begin{eqnarray}
\Psi_{\mathrm{p},k}(r,\varphi,z,t)=&&\Big(\frac{w_0}{w(z)}\Big)^{n+p+1}e^{ik(z-c t)}r^ne^{in\varphi}\label{lgaussbeamn}\\ &&\times L_p^n\Big(\frac{r^2}{w_0^2(1+i\frac{z}{z_R})}\Big)\exp\Big[-\frac{r^2}{w(z)^2}\nonumber\\ &&+i\frac{kr^2 }{2 R(z)}-i(n+p+1)\psi_G(z)\Big].\nonumber
\end{eqnarray}

The corresponding non-paraxial solution of the full d'Alembert equation is again obtained through the formula~(\ref{pbgfnp}).

\subsection{$\gamma$ beam}\label{gammab}

A different kind of a beam that can be obtained in a similar manner is represented by the incomplete gamma function. We devote a bit more space to it, since in some aspects this beam, not exhibiting Gaussian behavior in the perpendicular plane, differs from any other and up to our knowledge seems not to have been dealt with in the literature. The defining expression is found by plugging
\begin{equation}
\beta_n(s)=s^{n-2},\;\;\;\; \mathrm{where}\;\; n = 2, 3, \ldots,
\label{gamma}
\end{equation}
into~(\ref{gaussb}) and (\ref{inp}). After completing calculations equivalent to those of the earlier examples, this choice yields the following result
\begin{equation}
\psi(r,\varphi,z)=2^{n-1} \frac{1}{r^n}e^{in\varphi}\gamma\Big(n,\frac{r^2}{\alpha(z)}\Big),
\label{gammabeam}
\end{equation}
where $\gamma(n,z)$ stands for the incomplete gamma function. It is straightforward to verify that $\psi$ in this form satisfies the paraxial equation~(\ref{paraxial}). Using the known expansion~\cite{span}
\begin{equation}
\gamma(n,w)=(n-1)!\, e^{-w}\sum_{j=0}^\infty\frac{w^{j+n}}{(n+j)!},
\label{expg}
\end{equation}
the $\gamma$ beam may be written as a superposition of Gaussian beams and similar formula can be written in terms of LG beams as well.

The behaviour for large $r$ is now hidden in the asymptotics of the function $\gamma(n,w)$, which reads~\cite{span}
\begin{equation}
\gamma(n,w)\sim \Gamma(n)-w^{n-1} e^{-w},\;\;\;\; \mathrm{as}\;\; |w|\rightarrow\infty.
\label{asg}
\end{equation}
Unlike other examples of this section, the profile turns out not to be Gaussian since for $r\rightarrow\infty$, one obtains
\begin{equation}
\psi(r,\varphi,z)\sim 2^{n-1} \frac{(n-1)!}{r^n}e^{in\varphi},
\label{gammabeamas}
\end{equation}
Due to the condition $n>1$ it is, however, still square-integrable. The non-Gaussian character is unique among the considered examples and stems from the presence of the special value $-2$ in the power of $s$ in~(\ref{gamma}). 

\begin{figure}[h]
\begin{center}
\includegraphics[width=0.50\textwidth,angle=0]{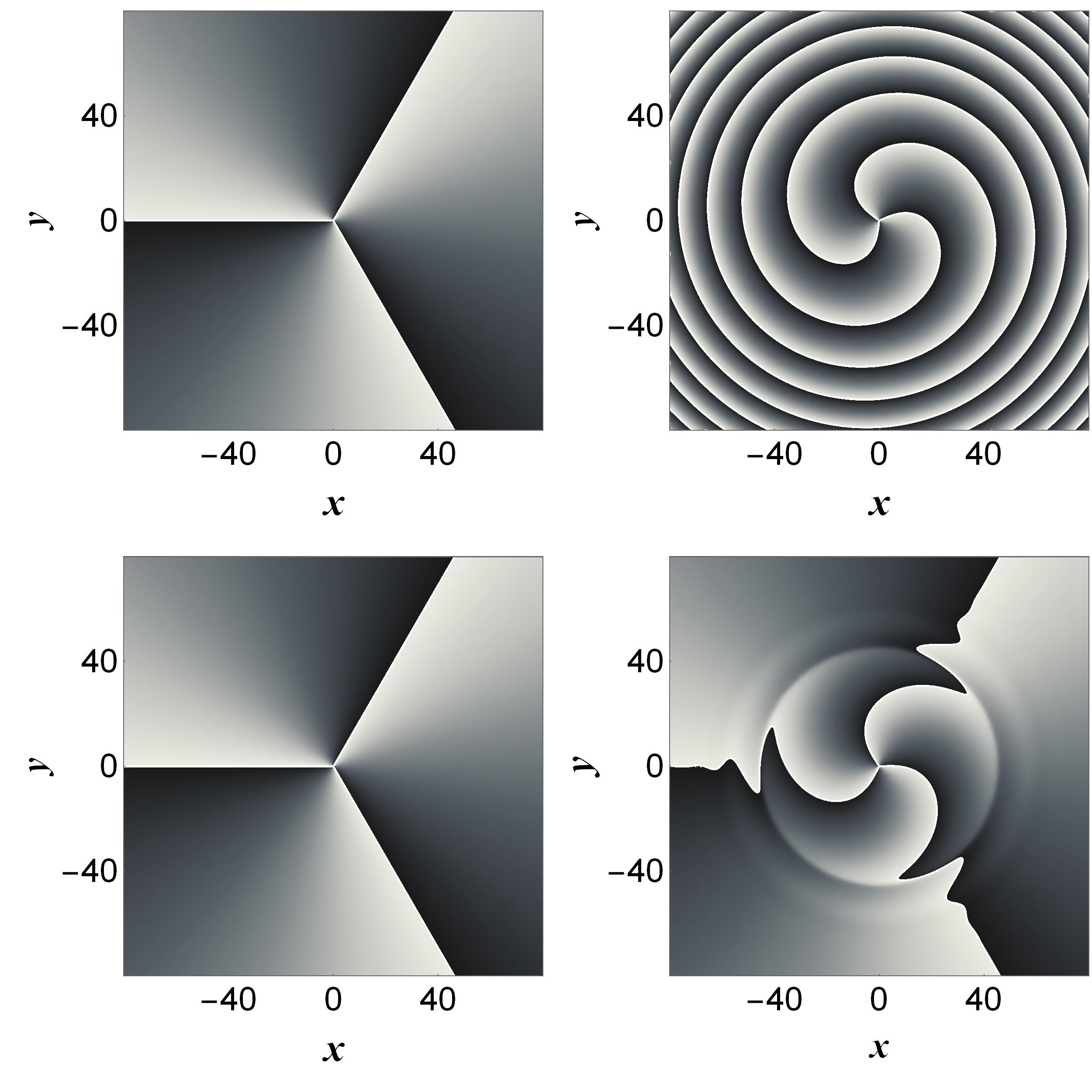}
\end{center}
\caption{The phases of the Gauss beam (upper plots) and $\gamma$ beam (lower plot) depicted in two planes: $z=0$ and $z=100$. The quantities on axes are measured in $k^{-1}=\lambda/2\pi$. The value of the phase, modulo $2\pi$, is represented continuously by means of the grayscale from $-\pi$ (black color) to $\pi$ (white color). The spiral is typical for all the Gaussian-type beams because of  the interplay between phase factors $e^{in\varphi}$ and $e^{ikr^2/R(z)}$. The additional rotation of the entire picture with increasing $z$ appears if the factor $e^{ikz}$ is taken into account.}
\label{phases}
\end{figure}

It should be noted that the phase of~(\ref{gammabeamas}) is now $r$-independent. Therefore, in each plane $z=\mathrm{const}$, lines of constant phase are (asymptotically) radially oriented unlike all other beams dealt with earlier that display spiral character. This is depicted on Fig.~\ref{phases}, where phases of the Gaussian beam~(\ref{gbnp}) and of the $\gamma$ beam~(\ref{gammabeam}) are compared for $n=3$ in two planes: $\zeta=0$ and $\zeta=100$. The spiral pattern with growing $\xi$ (i.e., $r$) was due to the presence of the factor $e^{ikr^2/R(z)}$ in all other beams. 

Naturally, while moving along the $z$ axis, the helical character typical for beams endowed with orbital angular momentum occurs due to the interplay between factors $e^{in\varphi}$ and $e^{ikz}$, the latter stemming from~(\ref{epi}).

For small $w$ one simply has~\cite{span}
\begin{equation}
\gamma(n,w)\sim \frac{w^n}{n}.
\label{smg}
\end{equation}
and the beam exhibits behavior typical for a vortex of order $n$:
\begin{equation}
\psi(r,\varphi,z)\sim \frac{2^{n-1}}{n} \frac{1}{\alpha(z)^n}\,r^ne^{in\varphi}.
\label{gammabeamam}
\end{equation}
The $\gamma$ beam in terms of conventional optical notation has the form
\begin{equation}
\Psi_{\mathrm{p},k}(r,\varphi,z,t)=e^{ik(z-c t)}\frac{1}{r^n}\,e^{in\varphi}\,\gamma\Big(n,\frac{r^2}{w_0^2(1+i\frac{z}{z_R})}\Big).
\label{pgam}
\end{equation}

\begin{figure}[h!]
\begin{center}
\includegraphics[width=0.50\textwidth,angle=0]{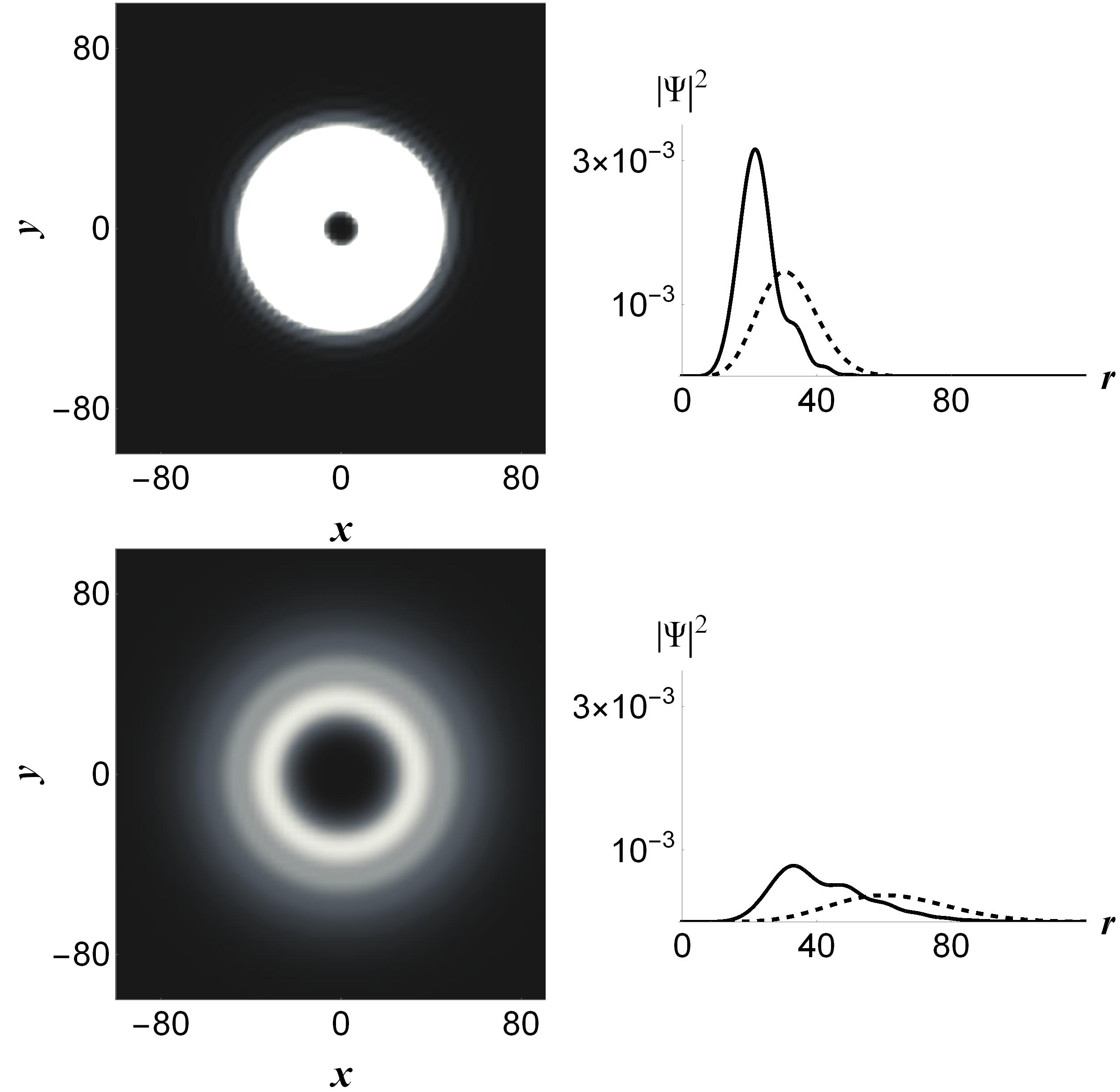}
\end{center}
\caption{The (dimensionless) intensity distribution of the $\gamma$ beam in the planes $z=50$ (upper drawings) and $z=100$ (lower drawings) for $n=3$, represented by the normalized value of $|\Psi(r,\varphi,z,t)|^2$. The distance $r$, as in Fig.~\ref{phases}, is measured in $k^{-1}$. Bright regions on the left represent those of high intensity. The right graphs display the same intensity in the radial direction (e.g. $|\Psi(r,0,z,t)|^2$) for $\gamma$ beam (solid lines) and for comparison for the Gaussian beam (dashed lines). The particular values on the ordinate are inessential, but can serve to compare the upper and lower graphs. On the left plot these values are rescaled so that black color represents the zero value, and white color the maximum value from the right plot.}
\label{int}
\end{figure}

In Fig.~\ref{int} the intensity of the $\gamma$ beam in the perpendicular plane is depicted in two planes: upper plots correspond to $z=50$ and lower ones to $z=100$. In addition, the radial plot (i.e. that of $|\Psi(r,0,z,t)|^2$) and its comparison with the corresponding curve for the Gaussian beam are shown on the right (naturally, despite the presence of $t$, these expressions are time independent). The expanding area of low intensity in the middle, represented by the dark disks on the left plots and by the corresponding minima close to the origin on the right ones, but with the relatively high barrier, might eventually serve as traps for neutral particles with negative polarizability such as blue-detuned atoms. It can be seen that the wave spreads as it moves along the $z$-axis, which is typical behavior for paraxial beams. At the waist, i.e., for $z=0$, the black disk becomes extremely narrow in the scale used and invisible in the figure. 

Substituting $z\longmapsto \zeta$ under the $\gamma$ function the solution of the non-paraxial equation can be immediately obtained, according to~(\ref{pbgfnp}), i.e.,
\begin{equation}
\Psi_{\mathrm{A},k}(r,\varphi,\zeta,\eta)=e^{ik\eta}\frac{1}{r^n}\,e^{in\varphi}\,\gamma\Big(n,\frac{r^2}{w_0^2(1+i\frac{\zeta}{z_R})}\Big).
\label{pgama}
\end{equation}

\subsection{Kummer-Gaussian (Hypergeometric-Gaussian) beam}\label{kummer}

As can be seen from the functions $\beta_n(s)$, three of the beams dealt with above (i.e. the Gaussian beam, the LG beam and the $\gamma$ beam) belong to the same class that can be defined through the universal form:
\begin{equation}
\beta_n(s)=s^{l-1}, \;\;\;\; \mathrm{where}\;\; l+n>-1.
\label{kum}
\end{equation}
The beam that emerges with this choice can be called the Kummer beam as it is expressed through the Kummer function. In order to obtain this result the same procedure is followed. We first write
\begin{equation}
\psi(r,\varphi,z)=e^{in\varphi}\int\limits_0^\infty \mathrm{d}s\, s^l J_n(r s) e^{-\alpha(z)s^2/4},
\label{phikum}
\end{equation}
and then performing the integration with respect to $s$, the universal envelope is found to be:
\begin{eqnarray}
\psi(r,\varphi,z)&&=\frac{2^l}{n!}\,\Gamma\Big(\frac{n+l+1}{2}\Big) \frac{1}{\alpha(z)^{(n+l+1)/2}}\label{phikumbeam}\\ &&\times r^n e^{in\varphi}\,{}_1F_1\Big(\frac{n-l+1}{2};n+1;\frac{r^2}{\alpha(z)}\Big)e^{-r^2/\alpha(z)},
\nonumber
\end{eqnarray}
where ${}_1F_1(a;b;w)$ is the Kummer (i.e. the confluent hypergeometric) function~\cite{span}. Setting $l=n+1$ the fundamental Gaussian beam is obtained, for $l=n+2p+1$ (\ref{phikumbeam}) represents the LG beam, and $l=n-1$ yields the $\gamma$ beam of the previous section.

In the standard notation~(\ref{phikumbeam}) takes the form
\begin{eqnarray}
\Psi_{\mathrm{p},k}(r, &&\varphi,z,t)=\Big(\frac{w_0}{w(z)}\Big)^{(n+l+1)/2}e^{ik(z-c t)}r^n e^{in\varphi}\nonumber\\ 
&&\times \,{}_1F_1\Big(\frac{n-l+1}{2};n+1;\frac{r^2}{\alpha(z)}\Big)\label{phikumbeams}\\
&&\times\exp\Big[-\frac{r^2}{w(z)^2}+ \frac{ikr^2 }{2 R(z)}-i\,\frac{n+l+1}{2}\psi_G(z)\Big],
\nonumber
\end{eqnarray}
where, conventionally, an overall constant has been omitted. The general non-paraxial mode is obtained as always with the substitution~(\ref{pbgfnp}). 

\section{Paraxial approximation}\label{parapp}

In the present section we would like to justify -- using the paraxial approximation -- the simple rule $\alpha(\zeta)\longleftrightarrow \alpha(z)$ which relates the d'Alembert solution to the paraxial one. 
Since a monochromatic function is now being looked for, the first step is to obtain from~(\ref{supa}) the Helmholtz solution via the integral~(\ref{helm}). The integration with respect to $t'$ yields the Dirac delta function:
\begin{eqnarray}
&&\Psi_{\mathrm{H},\omega/c}(r,\varphi,z,t)=2\pi e^{-i\omega t}e^{in\varphi}\int\limits_{-\infty}^\infty\mathrm{d}k e^{ikz}D(k)\label{hin}\\
&&\times\int\limits_0^\infty \mathrm{d}s\, s J_n(r s)\beta_n(s)e^{-w_0^2s^2/4}e^{-izs^2/4k}\delta\Big(\omega-ck-\frac{cs^2}{4k}\Big),
\nonumber
\end{eqnarray}
At this stage $\omega$ is still independent on $k$. To proceed further one has to incorporate the paraxiality condition $w_0\gg\lambda=2\pi c/\omega$  (or equivalently $z_R\gg\lambda$). Consequently the presence of the exponent $e^{-w_0^2s^2/4}$ constrains the domain, which contributes significantly to the integral~(\ref{hin}), to values of $s$ that are small relative to $1/w_0$. Consistently, considering the roots of the equation
\begin{equation}
h(k):=\omega-ck-\frac{cs^2}{4k}=0,
\label{req}
\end{equation}
which may be given the form
\begin{subequations}\label{roots}
\begin{align}
&k_+=\frac{1}{2}\Big(\frac{\omega}{c}+\sqrt{\frac{\omega^2}{c^2}-s^2}\Big)\approx \frac{\omega}{c}-\frac{s^2 c}{4\omega},\label{root1}\\
&k_-=\frac{1}{2}\Big(\frac{\omega}{c}-\sqrt{\frac{\omega^2}{c^2}-s^2}\Big)\approx \frac{s^2 c}{4\omega},\label{root2}
\end{align}
\end{subequations}
together with the well-known identity~\cite{de}
\begin{equation}
\delta(h(k))=\frac{\delta(k-k_+)}{|h'(k_+)|}+\frac{\delta(k-k_-)}{|h'(k_-)|},
\label{delta}
\end{equation}
it can be observed that the second term contains additional power of $\lambda^2/w_0^2$, since 
\begin{subequations}\label{ap}
\begin{align}
&\frac{1}{|h'(k_+)|}\approx 1,\label{ap1}\\
&\frac{1}{|h'(k_-)|}\approx c^2s^2/4\omega^2.\label{ap2}
\end{align}
\end{subequations}
Therefore, in the examined approximation only the contribution from $k_+$ is significant, and after integration over $k$ one gets
\begin{eqnarray}
\Psi_{\mathrm{p},\omega/c}&&(r,\varphi,z,t)={\cal C}\, e^{-i\omega t}e^{in\varphi}\int\limits_0^\infty \mathrm{d}s\, s J_n(r s)\beta_n(s)\nonumber\\
&&\times e^{iz\omega/c-izcs^2/4\omega}e^{-w_0^2s^2/4}e^{-izcs^2/4\omega}\label{fin}\\
=&&\;{\cal C}\, e^{i\omega(z/c- t)}e^{in\varphi}\int\limits_0^\infty \mathrm{d}s\, s J_n(r s)\beta_n(s)e^{-\alpha(z)s^2/4},
\nonumber
\end{eqnarray}
where $\cal{C}\,$ stands for an inessential overall constant. It is obvious that this expression is identical with~(\ref{inpf}), upon identification $\omega/c=k$. In conclusion, the simple substitution $\alpha(\zeta)\longmapsto \alpha(z)$ actually corresponds to the paraxial approximation. It allows to directly obtain a paraxial beam from a d'Alembert pulse (and conversely). This observation makes the method described in this work still more simple and more universal.

\section{Outlook}\label{sum}

The formulated method of the derivation of analytical formulas for arbitrary Gaussian-type beams exhibiting cylindrical symmetry seems to be fairly universal. It allows to obtain in a straightforward way practically all known beams of this type, both within the paraxial approximation and for the full d'Alembert equation. In particular, with this method a $\gamma$ beam is obtained whose transverse profile is not governed by the Gaussian function. 

It should be emphasized that the nearly trivial manner of implementing the paraxial approximation demonstrated in the work allows for a joint description of both paraxial beams and d'Alembert pulses.

Seemingly, the method can provide a starting point for generating expressions describing other beams as well, which, like $\gamma$ beam, do not necessarily have to possess a Gaussian character. This beam itself is worthy of detailed examination, which will be done elsewhere. In order to generate other beams it is sufficient that appropriate analytical formulas for Hankel transforms exist and that one ensures that the energy flux is finite. In particular, it seems feasible for the method to work also for some special truncated beams, although probably it will not be possible to find a suitable function $g_n(s)$ for instance for truncated Bessel beams at least as explicit formulas are concerned. For some specific cases where such analytical formulas cannot be found, the method can quite easily be used to describe numerically the required modes of radiation. \\

\begin{acknowledgments}
I express my thanks to Professor I. Bia{\l}ynicki-Birula for bringing work~\cite{ibbz} to my attention and for some advices on the d'Alembert equation.
\end{acknowledgments}

\end{document}